\documentclass[twocolumn,prb,showpacs]{revtex4}
\usepackage{graphicx}

\usepackage{dcolumn}
\usepackage{bm}
\makeatletter

\newcommand{\Rmnum}[1]{\expandafter\@slowromancap\romannumeral #1@}
\makeatother

\begin{document}

\title{Emergence of ferroelectric switching in self organized Pr$_{0.67}$Ca$_{0.33}$MnO$_3$ nanocrystal array}

\author{Vinay Kumar Shukla}
\email{vkshukla@iitk.ac.in}
\affiliation{Department of Physics, Indian Institute of Technology, Kanpur, 208016, India}
\author{Soumik Mukhopadhyay}
\email{soumikm@iitk.ac.in}
\affiliation{Department of Physics, Indian Institute of Technology, Kanpur, 208016, India}

\begin{abstract}
We discuss the emergence of low dimensional ferroelectricity in self-organized $Pr_{0.67}Ca_{0.33}MnO_3$ nanocrystalline arrays deposited on Si substrate. We observe sharp switching of electric polarization under external electric field with nearly rectangular hysteresis loops which persist up to 150 K. We discuss the polarization switching kinetics within the framework of Kolmogorov-Avrami-Ishibashi Model which gives maximum domain wall speed of $\sim$1 m/s. Polarized Raman spectroscopy shows evidence of isolated ferroelectric domains even at room temperature.
\end{abstract}

\maketitle

\section*{Introduction} 
The emergence of multiferroicity in general and ferroelectricity in particular at reduced dimensions have been at the forefront of research towards nanoelectronic applications in recent years~\cite{Ramesh,Yu, Sando, Boyuan, Lin, Bretos}. Besides the technological aspect, such studies have called into question our present understanding regarding the plausibility of the emerging functional properties at the nanoscale. For example, Lee \textit{et al.} recently proposed, generalizing on the theoretical calculations and experimental measurements on nanometer thick films of cubic perovskite SrTiO$_3$, that size reduction could in fact lead to emergence of ferroelectric properties in otherwise non-ferroelectric bulk systems~\cite{Lee}. Stengel \textit{et al.} demonstrated a novel mechanism of ferroelectricity at metal-oxide interface that led to an overall enhancement of the ferroelectric instability of the film, rather than its suppression~\cite{Stengel}. On the other hand, there are several studies which claim that below certain critical size, ferroelectricity disappears (for a review on this see Ref.~\cite{Shaw}). One of our own works~\cite{Shukla} regarding the direct experimental evidence of multiferroicity in charge ordered nanocrystalline manganite Pr$_{0.67}$Ca$_{0.33}$MnO$_3$ (PCMO), which discusses particle size driven tunability of ferroelectric polarization, belongs to the former category. The origin of ferroelectricity in PCMO in that case being linked with Zener polaron (ZP) ordering~\cite{Daoud}, eventually leads to suppression of ferroelectricity with reduction in particle size. However, ferroelectric response is enhanced in nanocrystalline samples so long as ZP ordering is not destroyed~\cite{Shukla}. Anyway, the finite conductivity and the associated problem of leakage current makes it difficult to get saturated P-E loops with sharp reproducible switching for such nanocrystalline PCMO systems which exhibit ZP ordering. In this article, we discuss the origin of sharp ferroelectric switching in self-organized arrays of PCMO nanoislands deposited on Si substrate.

Depending on the interface energies and the lattice mismatch, the hetero-epitaxial growth modes responsible for self organization of arrays of three dimensional islands are generally divided into two distinct categories: first, Volmer-Weber (VW) which is characterized by islands growth and second, Stranski-Krastanow (SK), which is characterized by layer by layer growth followed by islands growth~\cite{Shchukin}. There are several reports which utilize these growth modes to fabricate self-organized arrays of ferroelectric capacitors. For example, the formation of self organized ferroelectric triangular shaped nano-islands at the initial stage of growth in PbTiO$_3$ and PZT films grown on Pt(111)/SiO$_2$/Si(100) has been reported~\cite{Masaru, Nonomura}. M. Alexe \textit{et al.}~\cite{Alexe} studied the switching properties of self assembled Bi$_4$Ti$_3$O$_{12}$ ferroelectric memory cells by scanning force microscopy working in piezoresponse mode. On the other hand, so far as manganites are concerned, the focus has been more on growing self-assembled epitaxial films on Si substrate with SrTiO$_3$ buffer layer~\cite{Hunter, Pradhan}.

\section*{Experimental details}
 Thin films of PCMO were grown on Si(100) and $SrTiO_3$ (100) (STO) substrates by Pulsed Laser Deposition (PLD). The target of PCMO was prepared in two steps-  firstly, we have synthesized nanocrystalline sample by standard sol-gel method.  Powders of $Pr_6O_{11}$, $CaCO_3$ and $MnO_2$ having purity 99.99 $\%$  (Sigma Aldrich), used as starting materials were preheated to remove any absorbed moisture. The corresponding oxides were converted into nitrates by using 45-50 $\%$ concentrated nitric acid. Suitable amounts of citric acid and water were added to the mixed solution. This  solution  was  stirred for 1 hour and then slowly  evaporated  in  a water bath at temperature of 55-60 $^{\circ}$C until a gel was formed. The obtained gel turns into black porous powder at higher temperatures. The nanocrystalline sample was prepared by annealing the pelletized powder at 1000 $^{\circ}$C for 6 hours. Secondly, nanocrystalline sample was further sintered at 1400 $^{\circ}$C for 36 hours to obtain the PCMO target of diameter around 20 mm. Before the deposition of the thin films, silicon substrates (un-doped) were cleaned by Piranha solution to remove any organic residues. The substrates were further ultrasonicated for 15 minutes each in three different solutions- acetone, iso-propyl alcohol and distilled water. Finally the substrates were dried by nitrogen air gun and heated at 100 $^{\circ}$C for 15 minutes. 
     The excimer laser of wavelength $\lambda$ = 248 nm having energy density of 2.5 $J/cm^2$ was used during the deposition.  The target to substrate distance, pulse repetition rate, substrate temperature were 4.5 cm, 10 Hz and 500 $^{\circ}$C respectively. The films were deposited at two different oxygen pressures of 250 mTorr and 15 mTorr followed by in-situ annealing at 500 $^{\circ}$C for 15 minutes at an oxygen flow of 1 bar. Finally the films were subjected to post annealing in air at 800 $^{\circ}$C for 2 hours. The deposition time was kept at 40 minutes in all the samples. 
   
	The characterization of thin films and target was done by XRD $\theta$-2$\theta$ scans at room temperature by using PAN alytical Xpert diffractometer having wavelength of 1.54 {\AA} (Cu-$K_{\alpha}$) with scan rate of 1 $^{\circ}$ per minute. The surface topography of all the films were studied by atomic force microscopy (AFM). The magnetic properties were measured using Quantum design Physical Property Measurement System (PPMS) while the ferroelectric properties were studied by Radiant Ferroelectric tester (Precision Premier-II) which uses virtual ground circuitry. The dielectric constant and ferroelectric polarization measurements were carried out using ``Positive-Up-Negative-Down'' (PUND) method. A number of parallel electrodes of dimension 3 mm x 0.5 mm with separation $\sim$ 1 mm were made by silver paint (Alfa Aesar) in lateral (in-plane) contact geometry. The dimension of all the samples were 10 mm x 5 mm. The Polarized Raman Spectroscopy was carried out using STR Raman Spectrometer (Airix Corporation, Japan) having half wave plate as polarizer and analyzers of 0 degree and 90 degree.

\section*{Results and Discussion}

\subsection*{Surface topography and magnetic characterization}
\begin{figure*}
\includegraphics[width=8.3 cm]{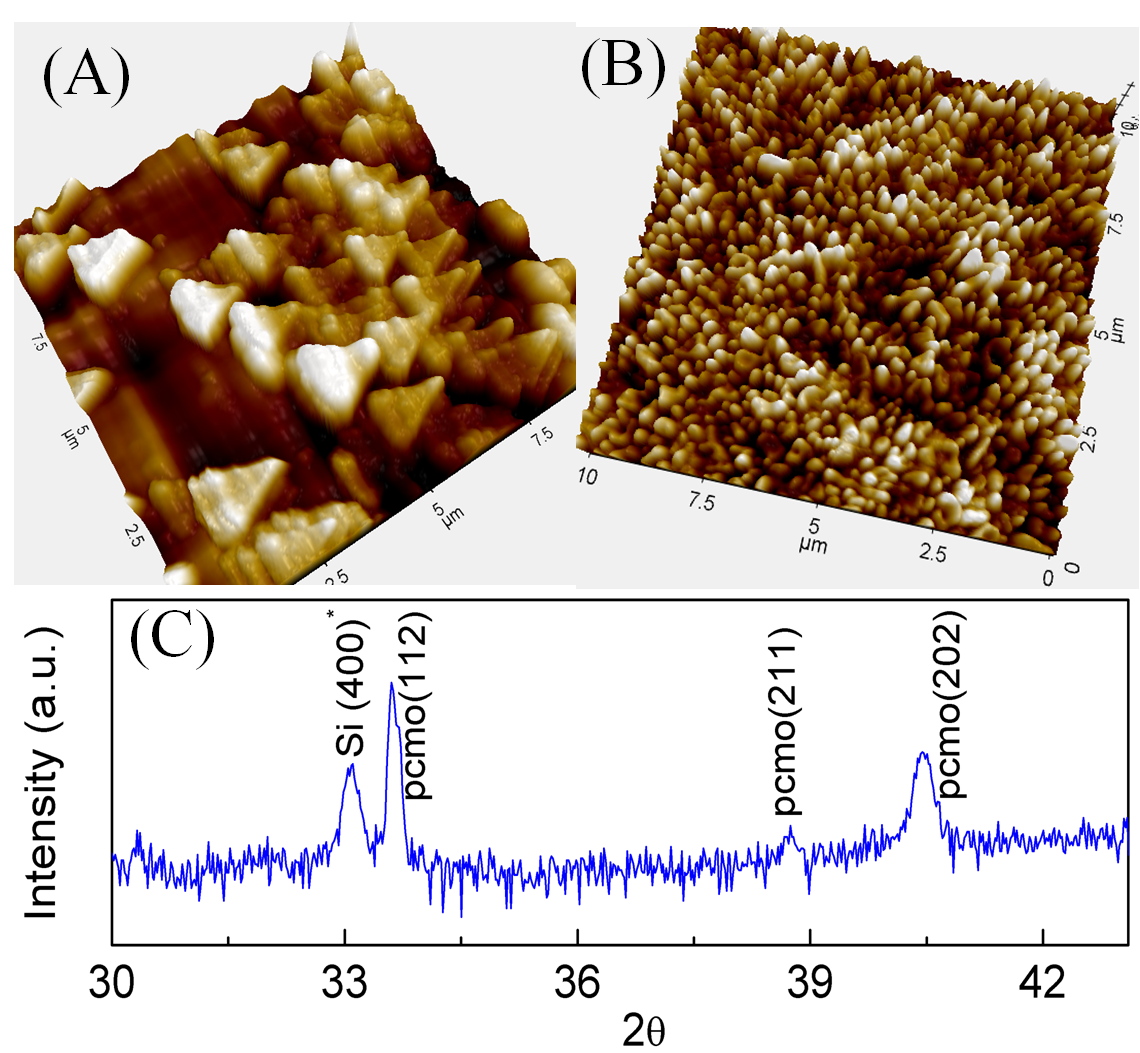}
\includegraphics[width=9 cm]{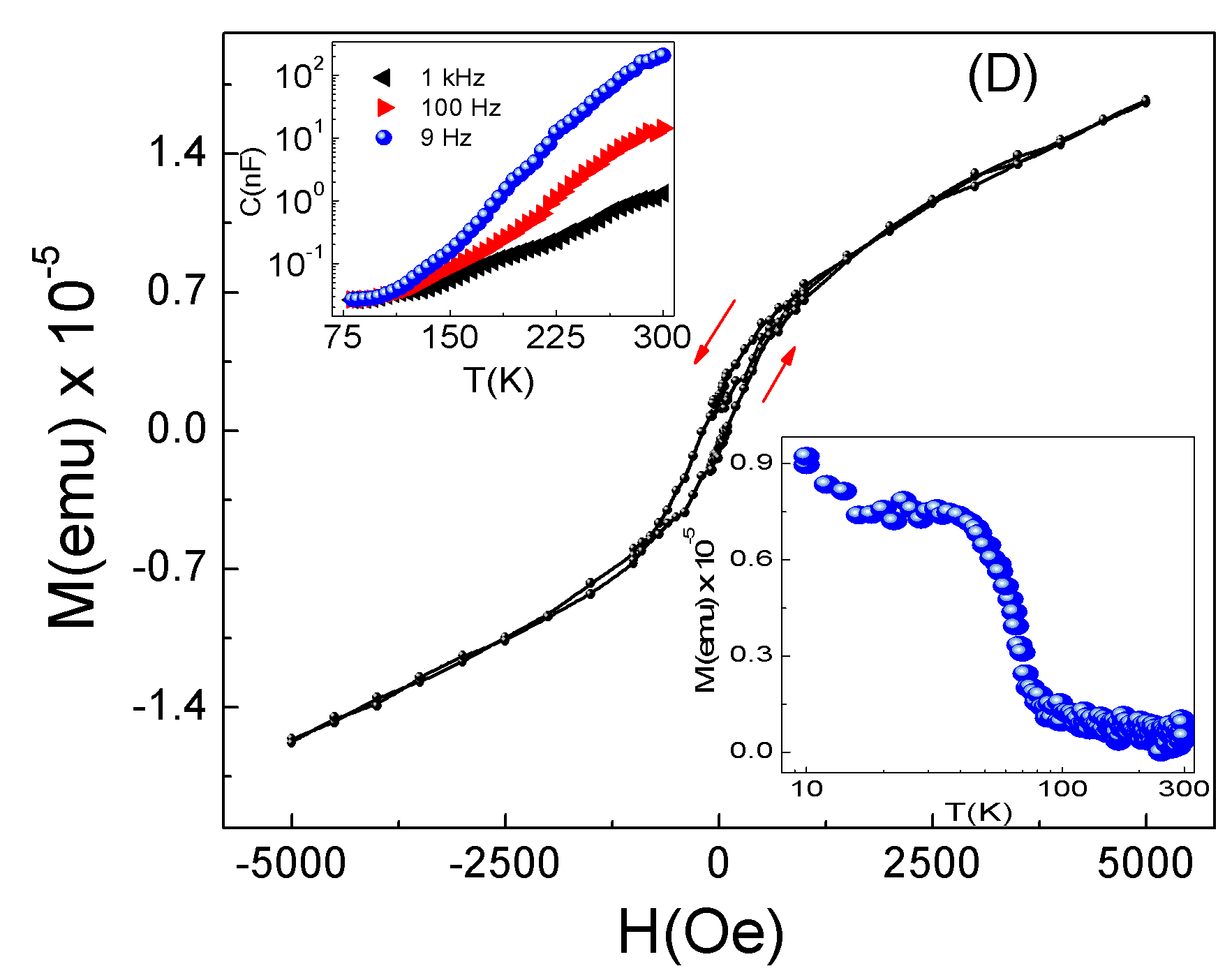}
\centering
\caption{Atomic force microscopy (AFM) images of PCMO$\vert$Si 1 (A) , PCMO$\vert$STO 1 (B). (C) X-ray diffraction pattern of PCMO$\vert$Si 1 indicates the polycrystalline nature of the film. $Si(400)^*$ represent (400) Bragg reflection of half wavelength of Cu-$K_{\alpha}$ radiation (D) Magnetization versus magnetic field hysteresis loop at 10 K for PCMO$\vert$Si 1. Inset(top left)- Temperature dependence of capacitance measured by ``PUND'' method with maximum applied field of 50 V/cm at different pulse widths. Inset (bottom right)- Temperature dependece of field cooled (FC) dc magnetization at H = 1 kOe.}\label{fig:mag}
\end{figure*}

 We label the PCMO films grown on Si(100) substrate at oxygen pressure of 15 mTorr as PCMO$\vert$Si 1 and the one at 250 mTorr as PCMO$\vert$Si 2. Similarly, PCMO films grown on STO at oxygen pressure of 15 mTorr and 250 mTorr are labeled as PCMO$\vert$STO 1 and PCMO$\vert$STO 2, respectively. The surface topography of PCMO thin films were examined by atomic force microscopy (AFM). The estimated maximum thickness of the films is around 100 nm. Fig.~\ref{fig:mag}A shows the AFM image of PCMO$\vert$Si 1 thin film indicating the formation of self assembled nano-islands with triangular shape anisotropy. The pseudo-cubic lattice parameter of polycrystalline PCMO is 3.83 {\AA}  whereas lattice constant of silicon is 5.43 {\AA} leading to huge lattice mismatch of $\sim 29.4 \%$. As a result, the x-ray diffraction pattern of PCMO$\vert$Si 1 (Fig.~\ref{fig:mag}C) shows (202) and (211) peaks apart from (112) peak with corresponding pseudocubic lattice constant of 3.41 {\AA}  rather than any epitaxial growth. The island formation in PCMO$\vert$Si 1 is facilitated by low substrate temperature of $500^{\circ}$C and low oxygen pressure during the deposition. The shape of the islands depends on the lattice structure at the surface of the underlying substrate. The crystal structure of Si(100) has 3-fold rotational symmetry ~\cite{Blochl} on the surface resulting into triangular shaped island growth of PCMO film. Similar type of self assembled Si/Ge nano structures had been observed at the surface of Si(111)~\cite{Bert, Kohler}. In general, with increasing oxygen pressure, the size of the plume increases, which in turn affects the thickness and surface morphology of the films.~\cite{Tselev} Consequently, for PCMO$\vert$Si 2 (fig not shown), density of the islands is increased and the shape anisotropy is reduced with enhanced oxygen pressure. Conversely, in lattice matched PCMO$\vert$STO 1 (Fig.~\ref{fig:mag}B) and PCMO$\vert$STO 2, epitaxial film growth is observed. The estimated rms roughness for PCMO$\vert$Si 1 is 40 nm whereas the same for PCMO$\vert$STO 1 is of few nms, clearly indicating different growth modes in the two cases. In lattice mismatched systems, the atoms of the films are strongly coupled to each other than to the substrate, leading to the formation of islands described within the framework of Volmer-Weber (VW) growth mechanism (For a review see Ref.~\cite{Shchukin})

Fig.~\ref{fig:mag}D shows the magnetization versus magnetic field hysteresis loop for PCMO$\vert$Si 1 at 10 K. The ferromagnetic hysteresis is observed in the low magnetic field range below 2500 Oe with absence of saturation till 5000 Oe. The temperature dependence of field cooled dc magnetization (bottom right inset of Fig.~\ref{fig:mag}D) taken at H = 1 kOe clearly indicates the onset of ferromagnetic ordering near 50 K. The Curie-Weiss(CW) temperature obtained from  inverse susceptibility gives the value of 40 K which is lower than that of nanocrystalline PCMO~\cite{Shukla}. However, the magnetic anomaly associated with ZP ordering is not observed in this case. Previous experiments on epitaxial film also report absence of ZP anomaly in the magnetic and transport measurements~\cite{Lees}. The dielectric measurements were carried out by ``PUND'' method~\cite{Naganuma, Fukunaga} in which sample is subjected to rectangular voltage pulses of maximum field 50 V/cm at different pulse widths ranging from 1 Hz- 1 kHz. In top left inset of Fig.~\ref{fig:mag}D, the monotonic increase of capacitance  with temperature is shown for PCMO$\vert$Si 1 with absence of any dielectric anomaly. Similar observations were recorded for other samples too. This is in agreement with previous reports on nanocrystalline PCMO~\cite{Shukla}. Overall, the dielectric and magnetic measurements fail to detect any structural phase transition but the presence of ferromagnetic correlation in self assembled nanocrystalline array of PCMO with corresponding Curie-Weiss temperature of 40 K is established.

\subsection*{Ferroelectric Hysteresis and switching kinetics}
\begin{figure*}
\includegraphics[width=13.5 cm]{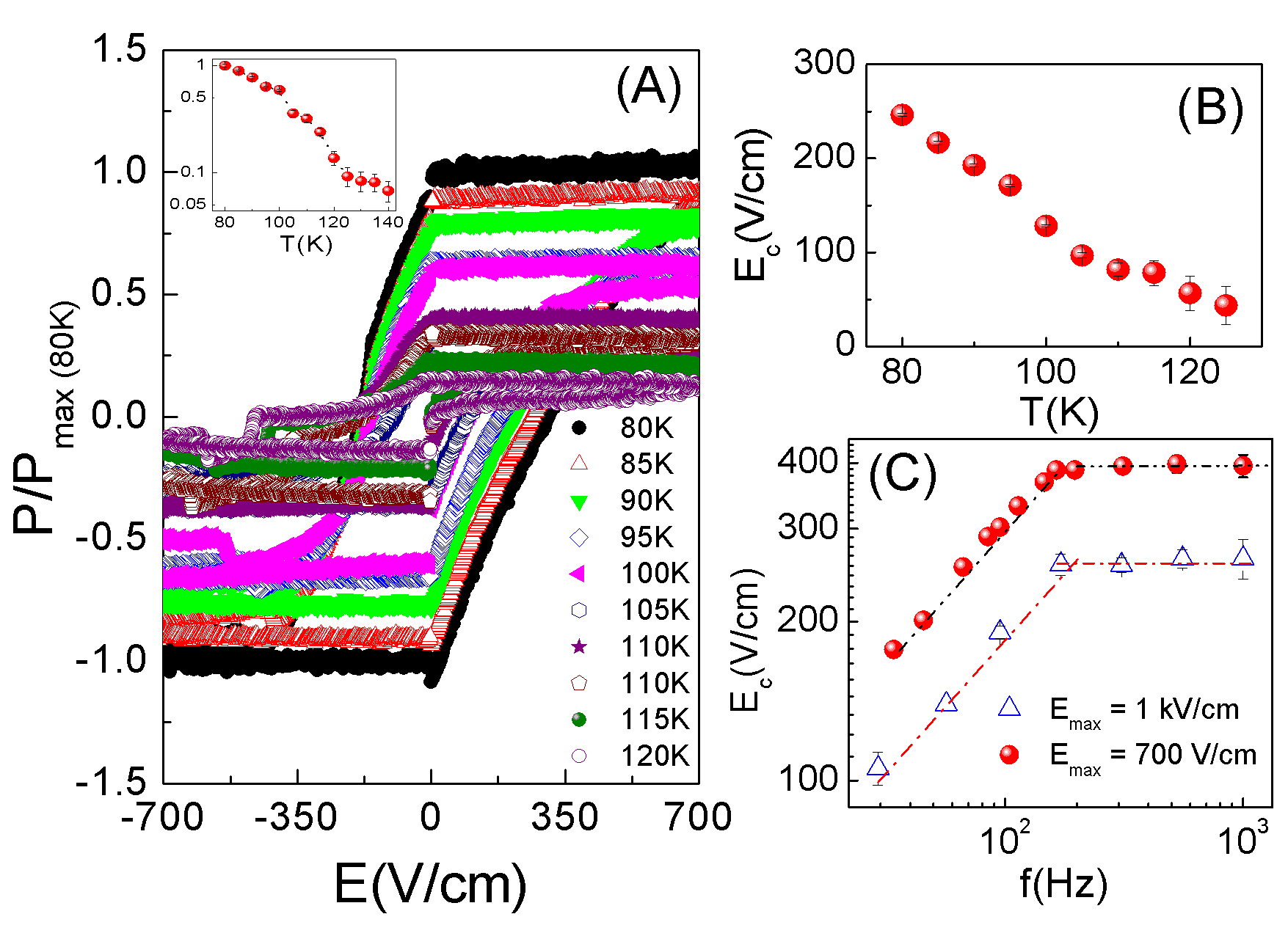}\\
\centering
\caption{(A) Remanent electric polarization of PCMO$\vert$Si 1 at different temperature from 80 K-120 K taken by ``PUND'' method at pulse width of 1 kHz with maximum applied field of 1 kV/cm. Inset: The corresponding saturation polarization versus temperature, (B) Corresponding temperature dependence of coercive field at pulse width of 1 kHz , (C) frequency dependence of coercive field at 80 K for two different maximum applied electric fields of 700 V/cm (solid red) and 1 kV/cm (open blue).}\label{fig:pund}
\end{figure*}

We have explored the ferroelectric properties by measuring remanent electric polarization based on PUND method. Traditional P-E loops are generally composed of ferroelectric, parasitic, or stray capacitance and conductive contributions. Therefore, PUND method or Double Wave method~\cite{Naganuma, Fukunaga} is a better tool to measure the switching polarization vs. electric field loops. Here, we have measured the remanent electric polarization using set of 5 `monopolar' triangular voltage pulses based on PUND method. First, the preset pulse sets the polarization of the sample in a certain direction and during which no measurement is made. Second, a switched triangular pulse of amplitude $V_0$ and pulse width $t_0$ is applied in the opposite direction of the preset pulse, which reverses the sign of polarization, and gives the half hysteresis loop associated with electric polarization which includes remanent and non-remanent components. In the third step, a `non-switched' pulse of same amplitude and pulse duration is applied in same direction as that of preceding pulse, which captures the half hysteresis loop containing only non-remanent contribution of polarization. The difference between half hysteresis loops measured in step 2 and 3 gives the switching polarization only. Similarly, sequence 4 and 5 are repeated but in opposite direction. The combination of half hysteresis loops obtained from the difference of sequence 2 and 3, 4 and 5 respectively, gives the complete four quadrant hysteresis $P_{SW}-E$ loops. Instead of rectangular pulses applied in capacitive measurements, here, we have applied triangular pulses so that rate of change of electric field over time should be fairly constant except for reversal points. The switching electric polarization versus electric field ($P_{SW}$-E) loops were studied for all four samples at different temperatures from 80 K-300 K. The maximum applied field was 1 kV/cm at different pulse widths ranging from 1 Hz-1 kHz. Fig.~\ref{fig:pund}A shows $P_{SW}$-E loops for PCMO$\vert$Si 1 at different temperatures from 80 K-120 K for pulse width of 1 Hz.  The saturation polarization decreases with increasing temperature (Inset of Fig.~\ref{fig:pund}A), typical of conventional ferroelectric materials. The observed $P_{SW}$-E loops shows sharp reproducible switching up to 150 K (Fig.~\ref{fig:pund}A). On the other hand, PCMO$\vert$Si 2, in which the coverage density of the islands is enhanced and shape anisotropy reduced, does not show ferroelectric hysteresis (fig not shown) within the same field range and temperature range of interest. This indicates that the ferroelectricity strongly depends on the film morphology.

\begin{figure}
\includegraphics[width=9 cm]{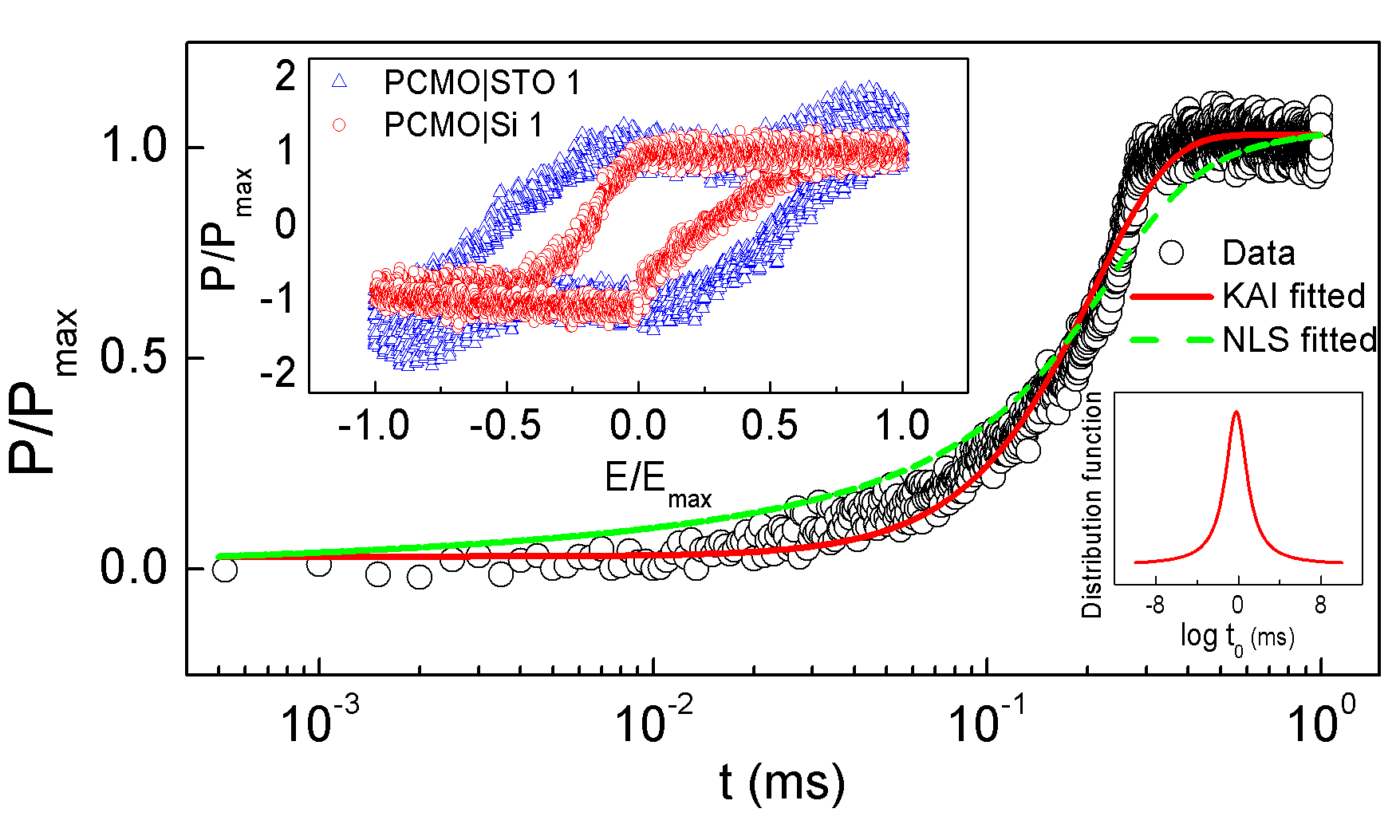}\\
\caption{Time dependence of normalized remanent electric polarization for PCMO$\vert$Si 1 for $E_{max}$= 600 V/cm  at 80 K. Solid Red and Dotted green lines represents fitting the experimental data by KAI and NLS model respectively. Inset (bottom right): shows the Lorentzian distribution for NLS fitted curve. Inset (top left): comparison between normalized P-E loops at 80 K for PCMO$\vert$Si 1 and PCMO$\vert$STO 1 at pulse width of 315 Hz with $E_{max}$ values of 1 kV/cm and 3 kV/cm respectively.}\label{fig:compare}
\end{figure}

Fig.~\ref{fig:pund}B shows the temperature dependence of coercive electric field ($E_c$) for PCMO$\vert$Si 1 at pulse width of 1 kHz. The maximum applied field ($E_{max}$) is sufficiently higher than the coercive field thus ruling out artifacts associated with P-E loops. The increase in coercivity with decreasing temperature is consistent with thermally activated ferroelectric switching. On the other hand, the frequency dependence of coercive field ($E_c$) at a constant temperature (Fig.~\ref{fig:pund}C) follows a power law behaviour~\cite{Scott} $E_c \sim f^\alpha $ with exponent $\alpha$ = 0.5.  It is evident from Fig.~\ref{fig:pund}C that the value of  $\alpha$  is nearly same for two different $E_{max}$ of 700 V/cm and 1 kV/cm indicating that $E_c$ values are fairly genuine. Moreover, if we compare PCMO$\vert$Si 1 and PCMO$\vert$STO 1 which have different growth modes, the coercivity in PCMO$\vert$STO 1 is almost 6 times than that of PCMO$\vert$Si 1 (Inset, Fig.~\ref{fig:compare}).

The switching time $t_s$, defined as time in which the value of polarization decreases to $90\%$ of its saturation value, ranges from $300{\sim}700 \mu$Sec depending on the applied electric field. The decrease of switching time with increasing $E_{max}$ indicates that the energy required for domain nucleation is mainly supplied by applied electric field. Moreover, since PCMO$\vert$Si 1 comprises of isolated self assembled nanostructures, the process of domain nucleation is most likely to occur separately in these islands and switching time captures the collective behaviour of fraction of volume switched. Although the Kolmogorov-Avrami-Ishibashi (KAI) model~\cite{Ishibashi, Yang} (which is based on the nucleation of domains expanding unrestricted under external electric field until it encounters another expanding domain) is extensively used to describe ferroelectric switching dynamics, it seems, ex ante, that nucleation-limited-switching (NLS) model~\cite{Tagantsev, Jo} which assumes the existence of multiple nucleation centers having independent switching kinetics would be the most suitable one to describe the system under study. The switching kinetics in the latter case is described in terms of the distribution function of the local nucleation probabilities. In order to clarify the nature of exponentially fast polarization switching, we have fitted the experimental data by KAI model as well as NLS model assuming a Lorentzian distribution of average logarithmic switching times (Fig.~\ref{fig:compare}). We find that even with the narrowest possible distribution (other than Delta function where NLS model reduces to KAI model) the KAI model gives a better fit to the experimental data. If P(t) is the switched remanent polarization at any instant t, its time evolution given by KAI model is,
\begin{equation}
P(t)= P_{max} [1-exp\{-(t/\tau)^n\}]
\end{equation}
where $P_{max}$ is the saturation polarization, $\tau$ is characteristic time of polarization reversal and n is the dimensionality over which the domain nucleation takes place.

\begin{figure}
\includegraphics[width=8.8 cm]{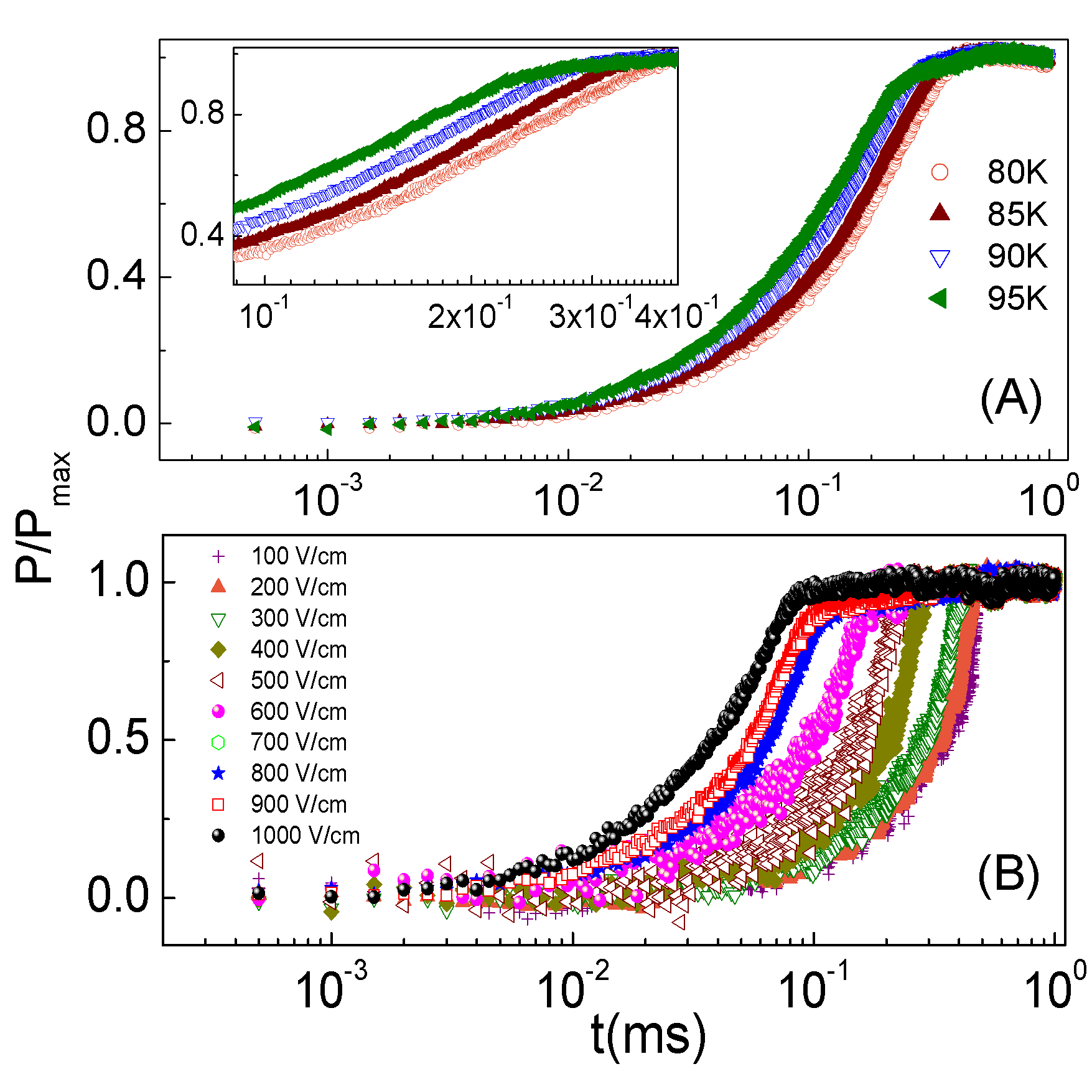}\\
\caption{Time dependence of normalized remanent electric polarization for PCMO$\vert$Si 1 (A) at different temperatures from 80 K-100 K at an applied electric field of 1 kV/cm; Inset: A blow-up of the same figure. (B) at 80 K for different applied electric fields from 100 V/cm - 1 kV/cm.}\label{fig:kai}
\end{figure}

 We performed switching time measurements by varying the maximum electric field $E_{max}$. At a given temperature, the electric field was increased to a set value $E_{max}$, at which we measured the elapsed time until the electric polarization is switched(Fig.~\ref{fig:kai}B). We fitted the results by the exponential function in KAI model~\cite{Ishibashi} to extract the characteristic timescale $\tau$. The characeteristic time clearly increases with decreasing temperature (Fig.~\ref{fig:kai}A) following the  Arrhenius behaviour $\tau= \tau_{0} exp(\epsilon_a/kT)$ (not shown) which implies thermally activated switching with corresponding time scale $\tau_0 = 10 \mu$Sec and energy scale $\epsilon_a = 20.4$ meV. Moreover, the characteristic time $\tau$ versus electric field obeys the Merz's Law which describes domain wall motion~\cite{Merz},
\begin{equation}
\tau =\tau_{\infty} exp({E_a}/E_{max})
\end{equation}
where $\tau_{\infty}$ is long time relaxation factor and $E_a$ is the corresponding activation electric field.Interestingly, the classic theoretical treatment of Merz's law developed by Miller and Weinreich~\cite{Miller} involves nucleation of an atomically thin triangular plate which expands in the same plane.

\begin{figure}
\includegraphics[width=8.4 cm]{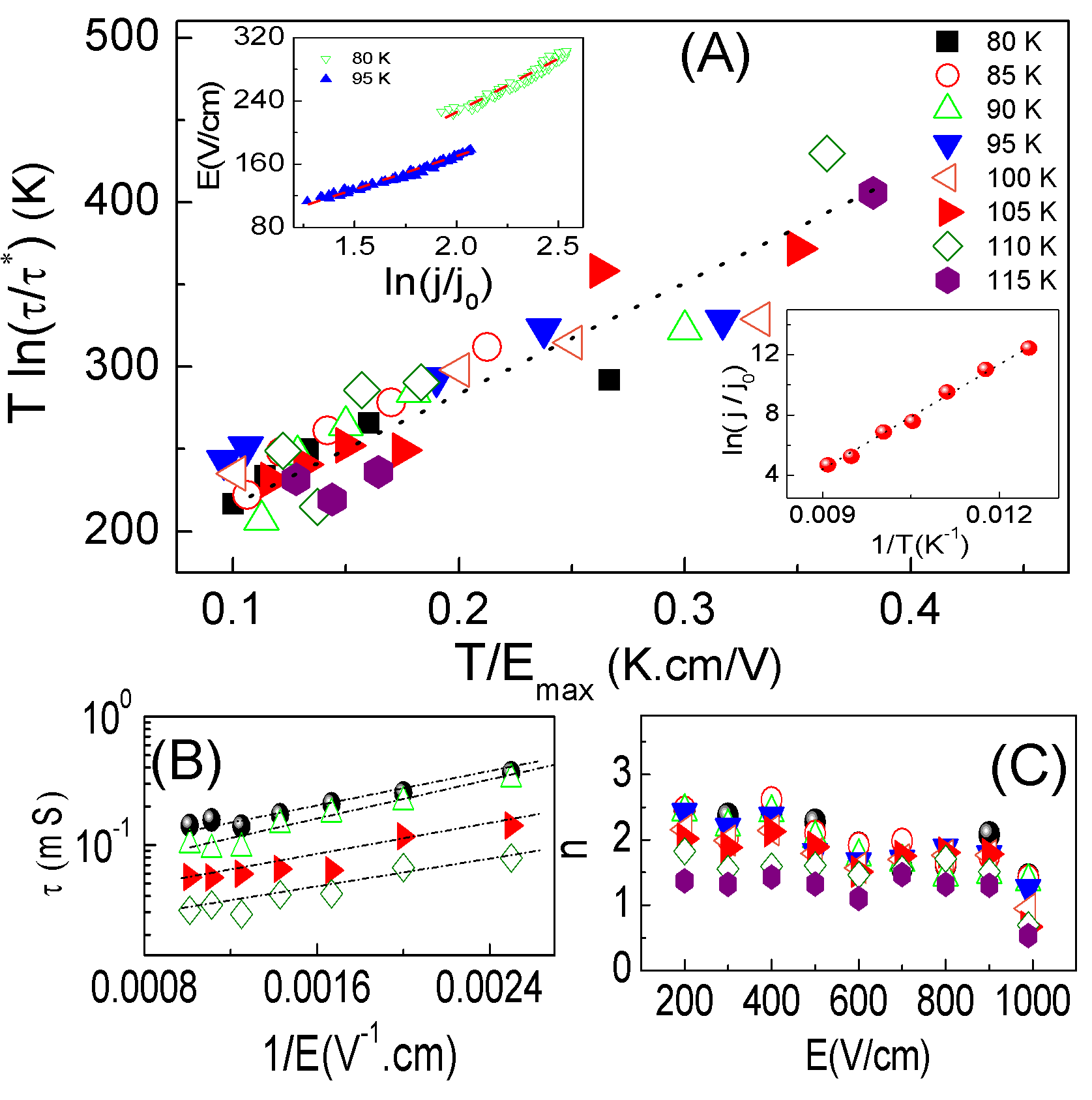}\\
\caption{A) Scaling plot of the switching time $\tau{(E_{max},T)}$ for several electric fields $E_{max}$ and temperatures where $T ln({\tau}/{\tau^{\ast}})$ is plotted against $T/E_{max}$. Top inset: Coercive field plotted as a function of logarithm of current density. Bottom inset: Logarithm of current density plotted against inverse of temperature. B) Characteristic time $\tau$ is plotted against inverse of electric field at different temperatures. C) The dimensionality n is plotted against electric field at different temperatures.}\label{fig:scaling}
\end{figure}

Fig.~\ref{fig:scaling}B shows the characteristic time $\tau$ plotted against inverse of applied electric field at different temperatures. The fitting parameters obtained from Merz's law are $E_a$ = 965 V/cm and $\tau_{\infty}$ as 10 $\mu$Sec. Apart from this, the electric field and temperature dependence of dimensionality is shown in Fig.~\ref{fig:scaling}C. Finally, characteristic time being a function of applied electric field $E_{max}$ as well as temperature $T$, we present our data of $\tau(E_{max}, T)$ in the form of a linear scaling plot of $Tln(\tau/\tau^{\ast})$ as a function of $T/E_{max}$, where $\tau^{\ast}$ is a scaling parameter. This makes interesting comparison with the ubiquitous scaling behaviour observed in case of nucleation and propagation mediated switching in nanomagnets~\cite{Wernsdorfer, Victora}. The slope of the power law scaling plot gives the effective activation field $E_a\sim$ 805 V/cm while the effective energy barrier calculated from the intercept turns out to be 24.2 meV. In order to calculate the activation volume, we follow the method prescribed by Chong \textit{et al.}~\cite{Chong}. The activation volume is computed at zero polarization i.e. near the coercive field in the P-E loop which corresponds to the fastest switching rate.

 Assuming that the direction of electric polarization is parallel to the applied electric field, the activation volume ($V^\ast$) can be calculated from the following relation~\cite{Chong}
\begin{equation}
\frac{\partial E}{\partial \ln \left({\frac{j}{j_0}}\right)}=\frac{K_{B}T}{V^\ast}
\end{equation}
where $j$ is the switched current density and $E$ is the electric field in the neighborhood of the coercive field $E_C$. One can also calculate the energy barrier from the slope of the
$\ln\left({\frac{j}{j_0}}\right)$ vs. $1/T$ plot. The energy barrier thus calculated is $25$ meV, close to the value obtained from the scaling plot, while the thermally activated volume (see Insets, Fig.~\ref{fig:scaling}A) turns out to be in the range $(4.4 nm)^{3}$ to $(7.3 nm)^{3}$ which is much smaller than the size of the individual islands. The activation volume is over-estimated since the nucleation field could be well below the coercive field. Anyway, the activation volume being much smaller than the size of the nanoislands reinforces our justification of the applicability of KAI model to the present case rather than NLS model. Taking the time scale associated with switching in separate islands to be $\sim 10-20 \mu$Sec  and the average size of the islands as $\sim 2-3\mu$m, one can estimate the maximum domain wall speed which is of the order of 1 m/s.

The values of n (dimensionality) obtained from fitting the normalized polarization versus time by KAI model~\cite{Ishibashi} varies between 1-2 depending on the applied electric field and temperature. The general observation is that n is close to 2 for lower electric field values and it decreases to 1 for higher electric fields. Moreover, the value of n decreases with increasing the temperature, too, suggesting a transition from predominantly isotropic domain growth to predominantly anisotropic growth with increase of the electric field and temperature. The decrease of n with increasing electric field is something unusual given the previous reports on other more conventional ferroelectrics. One of the possible reasons is that here the nucleation occurs simultaneously in various nano-islands. When the electric field is low, the domain nucleation occurs homogeneously in all the islands resulting in isotropic domain growth but as the electric field increases, the probability of nucleation among favorably oriented nano-islands increases compared to others leading to anisotropic domain growth. Similar phenomena is expected with increasing temperature as well.

\subsection*{Polarized raman spectroscopy}

\begin{figure}
\includegraphics[width=8.7 cm]{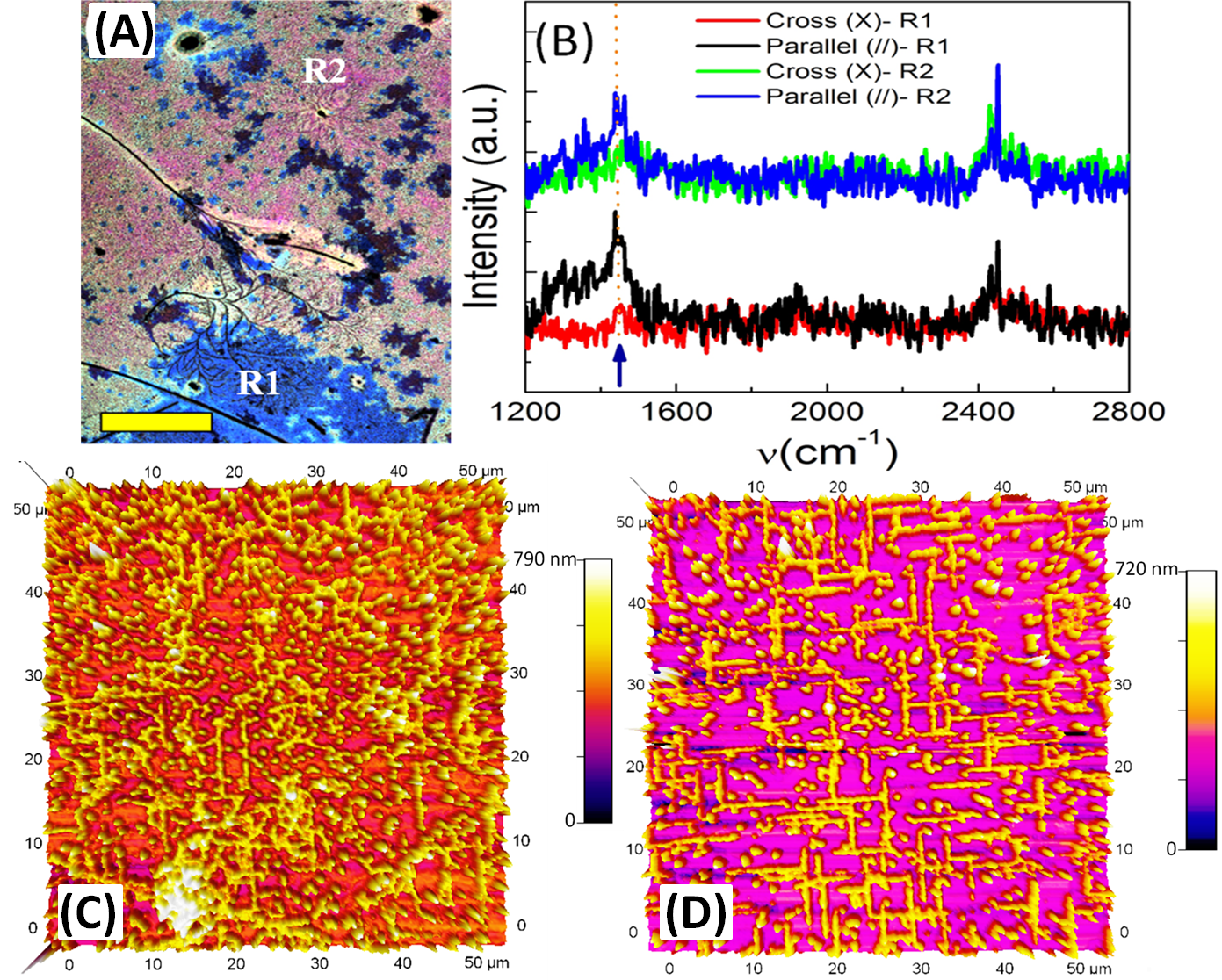}\\
\caption{(A) Optical microscopy image of PCMO$\vert$Si 1 shows different colour contrast of blue (R1) and yellow regions(R2). The scale bar represents 200 $\mu$m. (B) The room temperature Polarized Raman Spectra in parallel and crossed configurations of polarizer and analyzer are shown for R1 and R2 regions. (C) AFM image corresponding to the region R2 and (D) AFM image corresponding to the region R1 at an intermediate length scale showing region of different configurational symmetry of the array of nanocrystals.}\label{fig:raman}
\end{figure}
In order to explore the local inhomogeneity of the switching behavior, we have carried out polarized raman spectroscopy~\cite{Allemand} over different regions of the sample PCMO$\vert$Si 1. Fig.~\ref{fig:raman}A shows the optical microscope image of the sample at room temperature, which shows clear contrast between blue and yellow regions. Let us label the blue regions as R1 and yellow regions as R2. The polarized raman spectroscopy images are taken separately at the regions R1 and R2 (Fig.~\ref{fig:raman}B) of the sample in parallel and crossed configurations of the polarizer and analyzer. In region R1, the peak intensity at 1446 $cm^{-1}$ in crossed position is 28$\%$ as compared to that of parallel configuration, whereas in region R2, intensity only decreases to 90 $\%$ in crossed position compared to parallel position. The polarized raman spectroscopy conclusively identifies isolated regions like R2 showing ferroelectric raman modes surrounded by regions like R1 having only non-polar modes. The AFM images of PCMO$\vert$Si 1 taken at intermediate length scales between optical micrograph on one hand and Fig.~\ref{fig:mag}A on the other shows that the whole sample can be divided into two spatial regions with different densities of nanocrystals as well as having completely different configuration symmetries (Fig.~\ref{fig:raman}C, D). This might explain why we do not observe any net macroscopic polarization at room temperature. At room temperature there are isolated polar domains surrounded by non-polar regions which might undergo a ferroelectric phase transition at lower temperature. The minimum switching time scale observed in our case is 10 $\mu$sec at $E_{max}$ of 1 kV/cm, similar to $BiFeO_3$ thin films under much higher electric field of 250 kV/cm~\cite{Pantel}. On the other hand, the observed low coercive field ($\sim$ 200-400 V/cm), which is already underestimated considering the local inhomogeneity of the system, is similar to that reported for soft $BaTiO_3$ ceramics~\cite{Nagata} and some organic ferroelectrics~\cite{Sachio}.

\section*{Conclusion}

In conclusion, we have have presented evidence of emerging ferroelectricity in self organized PCMO nanostructured arrays grown on Si substrates. The polarization measurements reveal sharp ferroelectric switching up to 150 K in self assembled nanostructures with shape anisotropy, while ferromagnetic ordering is observed below 40 K. This is in sharp contrast with epitaxially grown films which show ferroelectric reversal at significantly higher electric field value. The polarization switching dynamics is attributed to the nucleation and growth of domains as described by KAI model occurring separately in different islands. The polarized raman spectroscopy along with optical and AFM imaging consistently provide evidence of local inhomogeneity and isolated ferroelectric domains even at room temperature. The observed fast switching of the electric polarization combined with low coercive field holds exciting prospects for future nano-electronic applications.

\section*{Acknowledgements}
VKS acknowledges CSIR, New Delhi for providing financial support. The authors wish to acknowledge The Department of Science and Technology, Govt. of India (FIST DST Project No. 20060268) for the Pulsed Laser Deposition facility.



\end{document}